
\magnification=1200
\baselineskip=18pt
\centerline{\bf COVARIANT PERTURBATIONS}
\centerline{\bf  OF DOMAIN WALLS IN CURVED SPACETIME}
\vskip2pc
\centerline {\bf Jemal Guven}
\vskip1pc
\it
\centerline {Instituto de Ciencias Nucleares }
\centerline {Universidad Nacional Aut\'onoma de M\'exico}
\centerline {A. Postal 70-543. 04510 M\'exico, D. F., MEXICO} \rm

\vskip2pc
\centerline{\bf Abstract}
\vskip1pc
{\leftskip=1.5cm\rightskip=1.5cm\smallskip\noindent
A manifestly covariant equation is derived to describe
the perturbations in a domain wall on a given background spacetime.
This generalizes recent work on domain walls
in Minkowski space and introduces a framework for examining the
stability of relativistic bubbles in curved spacetimes.
\smallskip}

\vskip2pc
\noindent{\bf I. INTRODUCTION}
\vskip1pc
Topologically stable field configurations
would have been produced
during any phase transitions that occurred in the early universe.
These could be monopoles, cosmic strings or domain walls depending on the
model. Even a
small abundance of these defects can have profound
cosmological consequences if they
were produced during or after inflation.
In practice, however,  calculations are often based
on the flimsy assumption that topological defects of
high symmetry are the ones which preponderate. Now, {\it perhaps} this can
be justified initially. At least in
Minkowski space, the semi-classical
approximation to quantum theory does appear to
predict an exponential suppression in the
materialization rate during a phase
transition of classical configurations with low
symmetry.[1] However, even then,
one cannot ignore the fact that there
will always be quantum fluctuations attending these
classical configurations.
If one is therefore to place any confidence in cosmological predictions
which rely on the assumption that symmetrical topological defects
maintain their shape in the course of their evolution,
such as the collapse of cosmic strings to form black holes, one had better
be sure no only that the evolution is stable but also that
a bound can be placed on the perturbation to prevent it from
substantially disrupting the process.[2]

In this paper, we
examine the evolution of small irregularities in a
domain wall propagating on a curved background spacetime.
This provides a generalization of the
work of Garriga and Vilenkin who examined the
stabily of domain walls in Minkowski space
and equatorial domain walls in de Sitter space.[3,4]
Because of the extra technical difficulty they involve,
we defer the examination of lower dimensional defects to a
subsequent publication.[5]

We begin in Section II with a
derivation of the equations of motion of the unperturbed
domain wall in the thin wall approximation.
We include a discussion of the boundary conditions which
must be implemented on any physical boundaries.
Our approach to perturbation theory in Section III
will be to expand the action describing domain walls in a
manifestly covariant way out to second order in the perturbation.
We work with the action directly rather than with
the equations of motion. This not only
possesses the advantage of preparing the ground for the
examination of the quantum theory of fluctuations, it also
facilitates the identification of appropriate boundary conditions.
In Section IV we discuss the equations of motion decribing
perturbations on various background spacetime
and domain wall geometries and compare our results with those
of Refs. [3] and [4].
We work in an arbitrary spacetime dimension, $N$.

\vskip2pc
\noindent{\bf II. THE EQUATIONS OF MOTION}
\vskip1pc

Let us consider an oriented domain wall $m$ in the thin wall
approximation. This is justified so long as the thickness of the
wall is much smaller than any other of its dimensions.
The wall is then
described by the timelike hypersurface
$$x^\mu=X^\mu(\xi^a)\,,\eqno(2.1)$$
$\mu=0,\cdots, N-1$, $a=0,\cdots, N-2$,
embedded in spacetime $M$ which we
describe by the metric $g_{\mu\nu}$.
The metric induced on the world-sheet of the domain wall is then
given by
$$\gamma_{ab}= X^\mu_{,a}
X^\nu_{,b}\,g_{\mu\nu}\,.\eqno(2.2)$$
The action which describes
the dynamics of this domain wall is given by
the Nambu form plus a possible spacetime volume term:
$$S[X^\mu,X^\mu_{,a}]=-\sigma\int_m d^{N-1}\xi \sqrt{-\gamma}
+\rho \int_{M_{int}} d^N x\sqrt{-g}\,.\eqno(2.3)$$
The first term represents the most simple generally covariant
action one can associate with the
wall, proportional to the area swept out by the
world sheet of the wall as it evolves.
The constant of proportionality $\sigma$ represents the energy density
of the wall in its rest  frame.

Before proceeding any further, we need to distinguish between
open domain walls possessing
a boundary, and closed ones which do not.
A closed domain wall need not be compact.
However, if it is not it must be infinite in all
directions. A spatial boundary at infinity can be ignored.

An oriented closed wall
provides a partition of spacetime into two regions, an interior
$M_{int}$ and an exterior $M_{ext}$ each supporting its own phase.
Neither region need be finite in spatial extent.
Let $\rho$ represent the energy density deficit in the region $M_{int}$.
In the description of the nucleation of bubbles, $M_{int}$ will be finite.
If $\rho$ is positive (negative), the
interior consists of true (false) vacuum.
We now associate an action with $M_{int}$ proportional
to the spacetime volume enclosed by the world sheet of the wall.
If this volume is infinite the associated action will be infinite.
However, the change in volume corresponding to a variation in the
embedding of compact support will always be finite. It will
not, therefore, affect the equations of motion.

If, on the other hand, the domain wall possesses a physical
spatial  boundary it clearly cannot
provide a partition of spacetime. The phases on either
side must coincide and it makes no sense to introduce the volume term.

In the derivation of the equations of motion
for the wall,
we need to distinguish between the physical boundaries
$\partial m_s$ of the world-sheet of an open wall
and the spacelike surface, $\partial m_t$,
we introduce to implement the variational
principle, and on which the initial and
final configurations of the domain wall are fixed.
One is forced to  impose appropriate boundary conditions on
the spatial boundary.

The equations of motion of the bubble wall are given by the
extrema of $S$ subject to variations
$X^\mu(\xi)\to X^\mu(\xi)+\delta X^\mu(\xi)$ which vanish on
$\partial m_t$:

$$-{\delta S\over \delta X^\mu} \equiv
\sigma\left[\Delta X^\mu + \Gamma^\mu_{\alpha\beta}(X^\nu)
\gamma^{ab} X^\alpha_{,a} X^\beta_{,b}\right]-\rho n^\mu =0\,,\eqno(2.4)$$
where $\Delta$ is the scalar Laplacian
$$\Delta ={1\over\sqrt{\gamma}}
\partial_a(\sqrt{\gamma}\gamma^{ab}\partial_b)\,,$$
$n^\mu$ is the unit normal to the worldsheet and
$\Gamma^\mu_{\alpha\beta}$ are the spacetime Christoffel symbols
evaluated on $m$. We comment on the derivation in the appendix.
If $\rho$ is zero,
Eq.(2.4) represents a higher dimensional generalization of the
geodesic equation describing the motion of a point defect.

Despite the nice analogy,
this form of the equations of motion is not very
useful in practice.  This is because all but one linear combination
of these equations are identically satisfied.
To see this, we note that, both on shell and off,
$$\nabla_b X^\mu_{,a} + \Gamma^\mu_{\alpha\beta}(X^\nu)
X^\alpha_{,a} X^\beta_{,b}=K_{ab} n^\mu
\,,$$
where $\nabla_a$ is the world-sheet covariant derivative
compatible with $\gamma_{ab}$ and
the extrinsic curvature tensor $K_{ab}$ is defined by [6]

$$K_{ab}=-X^\mu_{,a} X^\nu_{,b} n_{\mu;\nu}\,.\eqno(2.5)$$
The tangential projections
of the Euler-Lagrange derivatives of $S$ therefore vanish identically:
$$ {\delta S\over \delta X^\mu} X^\mu_{,b}=0\,.\eqno(2.6)$$
The geometrical reason for this redundancy
is the invariance of the action with respect to
world sheet  diffeomorphisms. In particular,
under the infinitesimal world sheet diffeomorphism,
$$\xi^a \to \xi^a+ \omega^a\,,$$
$\delta S =0$ which implies the (Bianchi) identities,
Eq.(2.6).

It is now clear that the
equations describing the world sheet are entirely equivalent to the single
equation
$$\sigma K=\rho\,.\eqno(2.7)$$
If $\rho$ vanishes, this is just the equation describing an extremal
hypersurface.
Thus Eq.(13) in Ref.[3] which was derived
for a domain wall propagating in Minkowski space
generalizes, with no surprises, to  a curved spacetime.
That there is only one independent equation describing the
dymanics of the domain wall is nothing to do with any
symmetry, spherical or otherwise, that the wall might possess.

If the wall possesses a boundary, we cannot justify
constraining the variation on this boundary so that we still have the
surface  term
$$\int_{\partial m_s} d^{N-2}u \sqrt{-f}
{\partial \sqrt{-\gamma}\over\partial X^\mu_{,a}}
l^a \delta X_\mu\eqno(2.8)$$
to contend with.
Here, we have parametrized the boundary $\xi^a=\Gamma^a (u^A)$, $A=0,\cdots,
N-2$ so that the metric which is induced on $\partial m_s$ is
$$f_{AB}= {\partial\Gamma^a\over \partial u^A}
{\partial\Gamma^b\over \partial u^B}\gamma_{ab}\,.$$
The inward pointing normal to $\partial_s m$ in $m$ is $-l^a$.
The boundary condition
$$X^\mu_{,a} l^a =0\eqno(2.9)$$
is sufficient to ensure that the boundary term vanishes. Physically, this
is the requirement that no momentum
be transferred across the spatial boundary.
Modulo the equations of motion, this will in turn
imply that the  boundary of the worldsheet must be a null surface.

\vskip2pc
\noindent{\bf III. THE QUADRATIC ACTION}
\vskip1pc

At lowest order, the dynamics of any irregularities in the
geometry of the wall
will still be described by the thin wall action  Eq.(2.3).
One might hope that the non-linearity of the equation of motion
would serve to damp out any irregularities which might
appear in the course of
the bubble's evolution in much the same way
as the non-linearity of the underlying field theory
is inclined to improve the thin wall approximation in certain models.[1]
However, the way it turns out (see for example in Refs.[3] and [4]),
is that sometimes it does but sometimes it does not.

One way to derive the equation of motion describing  the
perturbation in the wall is simply to consider the linearization
of  Eq.(2.7)
$$\delta K=0\eqno(3.1)$$
with respect to the displacement in the
worldsheet, $\delta X^\mu$. This is the method
exploited in Ref.[3] when the background is Minkowski space.
The approach we will follow,
will be to expand the
action out to quadratic order about
the classical solution satisfying Eq.(2.7).
Once this is done, it becomes a simple matter to obtain the
corresponding equations of motion. In addition, the variational
principle provides a guide to the implementation of
appropriate boundary conditions.

As we have seen, variations along tangential directions
correspond to world sheet diffeomorphisms.
The only diffeomorphism invariant
measure of the perturbation $\delta X^\mu$ in the wall is the scalar
$$\Phi \equiv n_\mu\delta X^\mu\eqno(3.2)$$
representing the normal projection of the
spacetime displacement vector $\delta X^\mu$. This single scalar will
now completely characterize the perturbation in the domain wall.

The simplest way to evaluate the
quadratic action is to
introduce Gaussian normal coordinates for spacetime
adapted to the world-sheet hypersurface. Thus,
from each spacetime point $P$ in the neighborhood of $m$, we
drop the geodesic from $P$ which
intersects $m$ orthogonally at the point $P^\prime$.
$P$ is then uniquely characterized by
the coordinates $\xi^a(P^\prime)$ and the
proper distance $\eta$ along the geodesic.
Permitting ourselves
an abuse of notation which should not lead to confusion, we
denote the non-trivial spacetime metric components by
$\gamma_{ab}(\eta,\xi^a)$.
We note that $\gamma_{ab}(0,\xi^a)=\gamma_{ab}(\xi^a)$ and that
$$K_{ab} =  {1\over 2}\gamma_{ab}^\prime\eqno(3.3)$$
where the prime denotes the proper derivative along the normal
and we evaluate it at $\eta=0$.
With respect to these coordinates, $\Phi$ is simply
the component of the variation $\delta X^\mu$ along the normal,
$\delta X^\eta$.
The first and second variations in $\gamma_{ab}$ are
$$\eqalign{\gamma_{ab}^{(1)}= &
2 K_{ab } \Phi \cr
\gamma_{ab}^{(2)}=& 2 K_{ab}^{\prime} \Phi^2
- 2 \Phi_{,a} \Phi_{,b}\,.\cr}\eqno(3.4)$$
To second order in $\Phi$,
$$\sqrt{- \gamma}^{(2)}={1\over 4}
\sqrt{-\gamma}\left[
\gamma^{ab}\gamma^{(2)}_{ab} +
{1\over 2}\left(\gamma^{(1)2} - 2\gamma^{(1)ab}\gamma^{(1)}_{ab}\right)
\right]\,.\eqno(3.5)$$
Dropping a divergence, and exploiting the constancy of $K$ implied by the
background equation of motion, the corresponding
second order action can be written

$$A^{(2)}=-{1\over2}\int_m d^D\xi \sqrt{- \gamma}
\left[\Phi \Delta \Phi  +
(K^\prime +K^2) \Phi^2  \right]\,.\eqno(3.6)$$
We must however be sure that we can discard the divergence with impunity.
This decomposes into surface terms on $\partial m_s$ and
$\partial m_t$.
The surface term on $\partial m_t$ causes no problem because the appropriate
boundary condition in the variational principle is
$\delta \Phi=0$. However,
to ensure the vanishing of the
boundary term on $\partial m_s$ we impose the Neumann boundary condition

$$ l^a\nabla_a \Phi=0\eqno(3.7)$$
there. This is the perturbative statement of Eq.(2.9) ensuring that the
perturbed surface remains null on any finite boundary.

This form of the second variation
is not very useful because it involves $K^\prime$. However, it is
simple to express $K^\prime$ in terms of
more familiar world-sheet and spacetime scalars. We note that
$K^\prime$ appears linearly in the spacetime scalar curvature,
$^N R$. The easiest way to eliminate $K^\prime$ is therefore to exploit the
Ricci identity
$$D_\mu D_\nu n^\alpha- D_\nu D_\mu n^\alpha
= ^N\! R_{\mu\nu\alpha\beta} n^\beta\,.$$
We contract on $\mu$ and $\alpha$ and project onto $n^\nu$:
$$n^\nu D_\mu D_\nu n^\mu- n\cdot D (D\cdot n)
= ^N\! R_{\mu\nu} n^\mu n^\nu\,.$$
We now rewrite the first term
$$n^\nu D_\mu D_\nu n^\mu =
D_\mu \left(n^\nu  D_\nu n^\mu\right) -
D_\mu n^\nu D_\nu n^\mu\,.$$
The divergence vanishes because the normal to a hypersurface is a
spacetime gradient: $n_\mu=\partial_\mu \eta$.
$$K^\prime = K_{ab} K^{ab} +^N R_{\mu\nu} n^\mu n^\nu\,.$$
An alternative way to eliminate
$K^\prime$ is to exploit the dynamical Einstein equation for $K$
in the initial value formulation of general relativity
with the replacement of proper time by $\eta$.[7]


We must also expand the enclosed volume
out to second order. We find (see appendix, Eq.(A2))
$$V^{(2)} ={1\over2} \int_m \sqrt{-\gamma} K \Phi^2 \,.\eqno(3.8)$$

We now add Eqs.(3.67) and (3.8)
and again exploit the background equation of motion
to cancel the $K^2$ term against the volume
contribution.
$$S^{(2)}={\sigma\over2} \int_m d^D\xi \sqrt{- \gamma}
\left[\Phi \Delta \Phi  +  \left(
^N\! R_{\mu\nu} n^\mu n^\nu  + K^{ab} K_{ab}
\right)\Phi^2  \right]\,.\eqno(3.9)$$
In the elimination of $K^\prime$ we introduced
the quadratic in the extrinsic curvature, $K_{ab}K^{ab}$.
We can, however,  eliminate this term from
$S^{(2)}$ in favor of curvature
scalars by exploiting the
contracted Gauss-Codazzi equation:
$$^N R_{\alpha\beta\mu\nu} h^{\alpha\mu} h^{\beta\nu}
=^{N-1}\! R + K^{ab} K_{ab}-K^2 \,,\eqno(3.10)$$
where the projection,
$h^{\alpha\beta}= g^{\alpha\beta}- n^\alpha n^\beta $.
We then get

$$S^{(2)}={\sigma\over2} \int_m d^D\xi \sqrt{- \gamma}
\left[\Phi \Delta \Phi  + \left(
^N R_{\mu\nu} n^\mu n^\nu + ^N\! R_{\alpha\beta\mu\nu}
h^{\alpha\mu} h^{\beta\nu}  - ^{N-1}\! R + K^2
\right)\Phi^2  \right]\,.\eqno(3.11)$$
Finally, we use Eq.(2.7) to
eliminate $K$ in favor of $\rho$ and $\sigma$ and
add the spacetime curvature terms

$$^N\! R_{\mu\nu} n^\mu n^\nu
+ ^N\! R_{\alpha\beta\mu\nu} h^{\alpha\mu} h^{\beta\nu}
=^N\! R_{\mu\nu} h^{\mu\nu}\,,$$
to obtain

$$S^{(2)}={\sigma\over2} \int_m d^D\xi \sqrt{- \gamma}
\left[\Phi \Delta \Phi  + \left(
^N R_{\mu\nu} h^{\mu\nu}  -
^{N-1}\! R + \left({\rho\over\sigma}\right)^2
\right)\Phi^2  \right]\,.\eqno(3.12)$$

\vskip2pc

\noindent{\bf IV. THE LINEARIZED EQUATIONS}
\vskip2pc
The equation of motion for small perturbations is now given by

$$ \Delta \Phi  + \left(^N\! R_{\mu\nu} h^{\mu\nu}  -
^{N-1}\! R +
 \left({\rho\over\sigma}\right)^2 \right)\Phi =0 \,.\eqno(4.1)$$
This is a scalar wave equation for $\Phi$ on the curved background
geometry of the world-sheet. If this geometry is
flat, $^N R^\mu{}_{\nu\alpha\beta}=0$ and
we reproduce Eq.(22) of  Ref.[3].
The wave equation then involves only the intrinsic geometry of the
worldsheet.

Eq.(4.1) is an unconventional Klein-Gordon equation in several respects.

\item{1.} We first note that the perturbation possesses a
tachyonic mass whenever $\rho \ne 0$. When $\rho=0$ this mass is zero.

\item{2.} There is a non-minimal coupling of the
perturbation to the scalar curvature of the background world-sheet
geometry. This coupling is universal and independent of the
dimension of the geometry. In particular, there is no privileged dimension
in which the coupling becomes conformal.

\item{3.} When the ambient spacetime geometry is not flat, the
perturbation couples to the tangential projection of the
Ricci tensor. This is the only explicit dependence of $\Phi$ on the
spacetime geometry. It is this feature which distinguishes
the perturbation theory we are considering
from a conventional field theory on the
curved spacetime described by the metric $\gamma_{ab}$.
If, however, the background is a vacuum solution to the
Einstein equations, this term vanishes. In particular, it
will vanish on Schwarzschild spacetime.
We note also that the perturbation does not couple to
the Weyl part of the background curvature.

Normally, we would interpret a tachyonic mass to signal an instability.
However, what is more significant is the effective mass given by
$$m^2=
^{N-1}\! R -^N R_{\mu\nu} h^{\mu\nu} -
 \left({\rho\over\sigma}\right)^2\eqno(4.2) $$
which might be positive depending on the value of the
the first two terms. This could depend on the topology of the
domain wall about which we are perturbing. However, even
a tachyonic effective does not always signal an instability.
This will be the case for perturbations about
an equatorial bubble in de Sitter space.[4] The
expansion of de Sitter space introduces a damping term
into the Laplacian which annuls the
destabilizing effect of a tachyonic effective mass.
Indeed, sometimes the notion of stability itself is ambiguous. An example
is provided in Ref.[3]
where the stability of true vacuum bubbles in Minkowski space is
discussed. The Fitzgerald-Lorentz contraction in the
perturbation detected by an inertial observer in Minkowski space
is sufficiently large to render physically divergent perturbations
of the wall apparently convergent.

In a vacuum background geometry with a cosmological constant $\Lambda$
(this need not be de Sitter space), the coupling to the
spacetime Ricci tensor simplifies.
The background Einstein equations then read

$$^N R_{\mu\nu} = {2\Lambda\over N-2} g_{\mu\nu}\,,\eqno(4.3)$$
so that Eq.(4.1) reduces to the form
$$ \Delta \Phi  + \left(
2\left({N-1\over N-2}\right)\Lambda   -  ^{N-1}\! R +
 \left({\rho\over\sigma}\right)^2 \right)\Phi =0 \,.\eqno(4.4)$$
Suppose, in particular, that the background is
de Sitter space. We can express $^N\!R$ in terms of the
Hubble parameter $H$
$$^N\! R = N(N-1) H^2 \,.\eqno(4.5)$$
Let us also suppose that $\rho=0$. We describe de Sitter space by
Friedman-Robertson-Walker (FRW) closed coordinates
$$ds^2=-dt^2 + H^{-2} \cosh ^2 (Ht) d\Omega^2_{N-1}\,,\eqno(4.6)$$
where
$d\Omega^2_{N}$ is the line element on a round $N$ sphere.
We now consider a
spherically symmetric domain wall.
In general, if the domain wall
is spherically symmetric, its world-sheet is
isometric to an $N-1$ dimensional FRW closed cosmology
described by the line element

$$ds^2=-d\tau^2 + a(\tau)^2 d\Omega^2_{N-2}\,,\eqno(4.7)$$
where $\tau$ represents the proper time recorded on a clock moving with
the wall.
The location of the wall at any time $\tau$ is
specified by the polar angle $\chi$ marking the position
the $N-2$ sphere of the wall is embedded on the $N-1$ sphere.
$$d\Omega^2_{N-1}= d\chi^2 + \sin^2\chi d\Omega^2_{N-2}\,.$$

There are two qualitatively different kinds of trajectory.[2]
One of these consists of trajectories
which begin with zero size at the pole $\chi=0$ and grow to a
maximum value before recollapsing. The other
is a bounce which consists of a bubble
originating on the equator ($\chi=\pi/2$)
contracting to a minimum value of $\chi$ before bouncing back to the
equator. There is a solution with $\chi=\pi/2$ representing a
domain wall which spans the equator. Such solutions can be
interpreted as bubbles which
materialize from nothing through quantum processes [2] and as such
are particularly interesting. The world-sheet is now an embedded
$N-1$ dimensional de Sitter space with the same Hubble paramter

$$^{N-1}\! R = (N-1)(N-2) H^2\,.$$
The equation describing small perturbations about this solution
is then given by
$$\Delta\Phi +(N-1)H^2\Phi=0\,.\eqno(4.8)$$
This reproduces the expression obtained in Ref.[3].
We stress, however, that the technique used in Ref.[3] to derive
Eq.(4.2) depended sensitively on the fact that the embedded
domain wall spanned the equator.

Even though the effective mass in Eq.(4.2)
is tachyonic, this does not appear to
be significant.

In general, the world-sheet of the bubble
will not be an embedded $N-1$ dimensional de Sitter space.
If the bubble is collapsing the scalar curvature
of its world-sheet will diverge. We recall that the scalar curvature
corresponding to the line element (4.7) is given by
$$^{N-1}\!R =2(N-2)\left[ {\ddot a\over a} +
(N-3)\left({\dot a^2\over a^2} +{1\over a^2}\right)\right]\,,$$
where the dot refers to a derivative with respect to $\tau$.
Both $\dot a$ and $\ddot a$ remain finite as $a\to 0$
At some point, therefore, the effective mass of the perturbations about the
collapsing bubble will be rendered real. It is also true,
however, that as we approach the final stages of collapse, the
thin wall approximation will break down.

\vskip1pc
\centerline{\bf V. CONCLUSIONS}
\vskip1pc
We have provided a framework for the examination of
perturbations of domain walls on a given spacetime background.

This analysis can be extended in at least two
different directions.

The first is the
treatment of perturbations on lower dimensional topological defects.[5]
When the co-dimension of the world-sheet is $r$, there will be
$r$ scalar fields describing the perturbation,
one for the projection of $\delta X^\mu$
onto each of the $r$ normal vectors $n^{(i)\,\mu}$.
What is more, the
equations we obtain will generally be coupled in a non-trivial way.
To derive the equations of motion,
we need to develop a different line of attack.
The reason for this is that
for co-dimension $r>1$, we can no longer
exploit a Gaussian system of
coordinates. The description of the extrinsic geometry
will, in general, be more complicated.
Now there will be $r$ extrinsic curvature tensors,
one for each $n^{(i)\,\mu}$:
$$K^{(i)}_{ab} =- X^\mu_{,a} X^\nu_{,b} n^{(j)}_{\mu;\nu}\,.$$
In addition,
so-called torsion terms of the form
$$T^{(i)(j)}_a = n^{(i)\,\mu} X^\nu_{,a} n^{(j)}_{\mu;\nu}$$
which vanish on a hypersurface now appear with a
vengeance.[6]

The weak point in our treatment of
perturbations is that the domain wall has not been treated as
a source for gravity.
This is a serious limitation.
If we are to place any confidence in perturbation theory,
we need to accommodate the back-reaction.
When this is done, the simple extremal form Eq.(2.7) gets replaced by
the Lanczos equations,
$$\Delta K_{ab}  = 8\pi G \sigma \gamma_{ab}\,,\eqno(5.1)$$
where $\Delta K_{ab}$ is the discontinuity suffered by $K_{ab}$
across the domain wall.[8] These equations are very different from
Eq.(2.7). If they do possess Eq.(2.7)
as their limit when the coupling to gravity
is turned off, this is not obvious.
What is more, whereas the solution of
Eq.(5.1) is relatively straightforward
when the domain wall is spherically symmetric[9], the
treatment of perturbations about such a wall
is far from trivial. For now,  the displacement in the
wall $\delta X^\mu$ will couple to
perturbations in the spacetime metric with the generation
of gravitational waves. This is currently being examined.

\vskip1pc
\centerline{\bf ACKNOWLEDGEMENTS}
\vskip1pc
I would like to thank Alexander Vilenkin for discussing his
work on the subject during his visit to The University of Mexico.
The treatment of the second variation of the action is a
simplified version, exploiting normal
coordinates of Niall \'O Murchadha's treatment of the
same variation in another context.
\vskip1pc
\centerline{\bf APPENDIX}
\vskip1pc

The extrema of $S$ with respect to variations
which vanish on the boundary satisfy the Euler
Lagrange equations
$$\sigma\left[{\partial\over\partial \xi^a}{\partial \sqrt{-\gamma}
\over \partial X^\mu_{,a}}-
{\partial \sqrt{-\gamma}\over \partial X^\mu}\right]
-\rho {\delta V\over\delta X^\mu} =0\,.\eqno(A1)$$
Now
$${\partial \gamma\over\partial\gamma_{ab}}=
\gamma \gamma^{ab}\,,$$
so that
$${\partial \sqrt{-\gamma}\over\partial X^\mu_{,a}}
=\sqrt{-\gamma}\gamma^{ab} g_{\beta\mu} X^\beta_{,b}\,,$$
and
$${\partial\over\partial \xi^a}{\partial \sqrt{-\gamma}
\over \partial X^\mu_{,a}}=
g_{\mu\beta}\partial_a(\sqrt{\gamma}\gamma^{ab} X^\beta_{,b})+
\sqrt{\gamma}\gamma^{ab} g_{\mu\beta,\alpha} X^\alpha_{,a}X^\beta_{b}\,.$$
We also have
$${\partial \sqrt{-\gamma}\over \partial X^\mu}
= {1\over 2}\sqrt{\gamma}\gamma^{bc}
g_{\alpha\beta,\mu}X^\alpha_{,a}X^\beta_{b}\,.$$
The first derivatives of the spacetime metric appear in the
combination $\Gamma^\mu_{\alpha\beta}$.
The term in square brackets Eq.(A1) reproduces the
corresponding term appearing in Eq.(2.4). To complete the
derivation of Eq.(2.4), we note that
under the variation $X^\mu\to X^\mu+\delta X^\mu$,
the volume transforms by

$${\delta V\over\delta X^\mu}= \sqrt{-\gamma}\, n^\mu \,,$$
or alternatively, in the notation of section III:

$$\delta V = \int d^{N-1}\xi \sqrt{-\gamma} \Phi \,.\eqno(A2)$$
In this form, it is clear that the second variation is given by
Eq.(3.7).

\vskip1pc
\centerline{\bf REFERENCES}
\vskip1pc
\item{1.} S. Coleman {\it The Uses of Instantons} in
The Whys of Subnuclear Physics,
ed. by A. Zichichi, (Plenum Press, New York, 1979)
\vskip1pc
\item{2.} R. Basu, A.H. Guth and A. Vilenkin
{\it Phys Rev} {\bf D44} (1991) 340
\vskip1pc
\item{3.} J. Garriga and A. Vilenkin {\it Phys. Rev} {\bf D44} (1991) 1007
\vskip1pc
\item{4.} J. Garriga and A. Vilenkin {\it Phys. Rev} {\bf D45} (1992) 3469
\vskip1pc
\vskip1pc
\item{5.} J. Guven {\it ICN preprint}, {\it ``Perturbations of
a topolological defect as a theory of coupled
scalar fields in curved space"} (1993)
\item{6.} Eisenhart {\it Riemannian Geometry} (Princeton Univ. Press,
Princeton, 1947);
M. Spivak {\it Introduction to Differential Geometry} Vol. IV
(Publish or Perish, Boston MA); K. Kucha\v r {\it J. Math. Phys}
{\bf 17} (1976) 777, 792, 801
\vskip1pc
\item{7.} N \'O Murchadha (1987) Unpublished Notes.
For a review of the initial value formulation of general relativity, see
J.W. York in {\it Gravitational Radiation} ed. by N. Deruelle and
T. Piran, (North Holland, 1982)
\vskip1pc
\item{8.} W. Israel {\it Il Nuovo Cimento} {\bf 44B} (1966) 1
\vskip1pc
\item{9.} See, for example,
S. Blau, E. Guendelman and A.H. Guth {\it Phys. Rev} {\bf D35}
(1987) 1747
\vskip1pc

\bye